\documentclass[review,number,sort&compress]{elsarticle}
\usepackage{fullpage}
\usepackage{graphicx}
\usepackage{amsmath}
\usepackage{amssymb}
\usepackage{wrapfig}
\usepackage{lineno}
\usepackage{amsfonts}
\usepackage{verbatim}
\usepackage[normalem]{ulem}
\newcommand{\twentysquare}{20 $\times$ 20 cm$^2$~}
\newcommand{\oneinchsquare}{2.54 cm $\times$ 2.54 cm~}
\newcommand{\halfinchsquare}{1.27 cm $\times$ 1.27 cm~}

\newcommand{\mcppmt}{microchannel-plate photomultiplier~}
\newcommand{\mcppmts}{microchannel-plate photomultipliers~}
\newcommand{\Mcppmts}{Microchannel-plate photomultipliers~}

\newcommand{\micron}{$\mu$m~}
\newcommand{\LAPPDTM}{LAPPD\textsuperscript{TM}~}

\begin{document}
\begin{frontmatter}

\title{Capacitively coupled pickup in MCP-based photodetectors using a conductive metallic anode}

\author[UC]{E. Angelico}
\author[UC]{T. Seiss}
\author[Incom]{B. Adams}
\author[UC]{A. Elagin}
\author[UC]{H. Frisch}
\author[UC]{E. Spieglan}
\address[UC]{Enrico Fermi Institute, University of Chicago, 5640 S Ellis Ave, Chicago, IL 60637}
\address[Incom]{Incom, Inc., 294 SouthBridge Rd, Charlton, Massachusetts 01507}

\begin{abstract}
 
We have designed and tested a robust \twentysquare thin metal film
internal anode capacitively coupled to an external array of signal
pads or micro-strips for use in fast microchannel plate photodetectors.
The internal anode, in this case a 10nm-thick NiCr film deposited on a
96\% pure Al$_2$O$_3$ 3mm-thick ceramic plate and connected to HV
ground, provides the return path for the electron cascade charge.  The
multi-channel pickup array consists of a printed-circuit card or
glass plate with metal signal pickups on one side and the signal ground
plane on the other. The pickup  can be put in close proximity to
the bottom outer surface of the sealed photodetector, with no
electrical connections through the photodetector hermetic vacuum
package other than a single ground connection to the internal
anode. Two pickup patterns were tested using a small commercial
MCP-PMT as the signal source: 1) parallel 50$\Omega$ 25-cm-long
micro-strips with an analog bandwidth of 1.5 GHz, and 2) a
\twentysquare array of 2-dimensional square `pads' with sides of
1.27 cm or 2.54 cm.  The rise-time of the fast input pulse is
maintained for both pickup patterns. For the pad pattern, we observe
80\% of the directly coupled amplitude.  For the strip pattern we
measure 34\% of the directly coupled amplitude on the central
strip of a broadened signal.  The physical decoupling of the
photodetector from the pickup pattern allows easy customization for
different applications while maintaining high analog bandwidth.
\end{abstract}


\end{frontmatter}

\section{Introduction}

\Mcppmts (MCP-PMT) are specialized vacuum photodetectors typically consisting of
a photocathode, an amplification section consisting of several planes
of glass micropores, and a segmented anode from which the amplified
pulses are detected~\cite{Wiza}. A hermetic package provides
internal vacuum, mechanical support, and electrical
connections. The photocathode is typically on the inside face of the
top window; the anode electrodes are integrated into the bottom plate of
the package.

Unlike conventional photomultipliers that focus the photoelectrons
from the photocathode onto a discrete chain of smaller dynodes, the
\mcppmt is planar with a uniform transverse structure. The
photocathode is proximity-focused onto the amplification section,
which in turn is proximity-focused onto the anode plane. This geometry
in principle enables excellent spatial resolution with a minimum scale
size of a pore diameter, and excellent time resolution, set by jitter
in the primary photoelectron time of `first strike' and jitter in the
first few interactions in the amplification electron
cascade~\cite{timing_paper}.  Large (\twentysquare) MCPs made from a
glass substrate with resistive and secondary-emitting layers formed
with Atomic Layer Deposition have demonstrated gains greater than
10$^7$, time resolutions measured in 10's of picoseconds, and spatial
resolutions on the order of 700 microns
~\cite{timing_paper,Ossy_JINST_2013}.

%
%
\begin{figure}[t]
\centering
\includegraphics[width = 0.98\columnwidth]{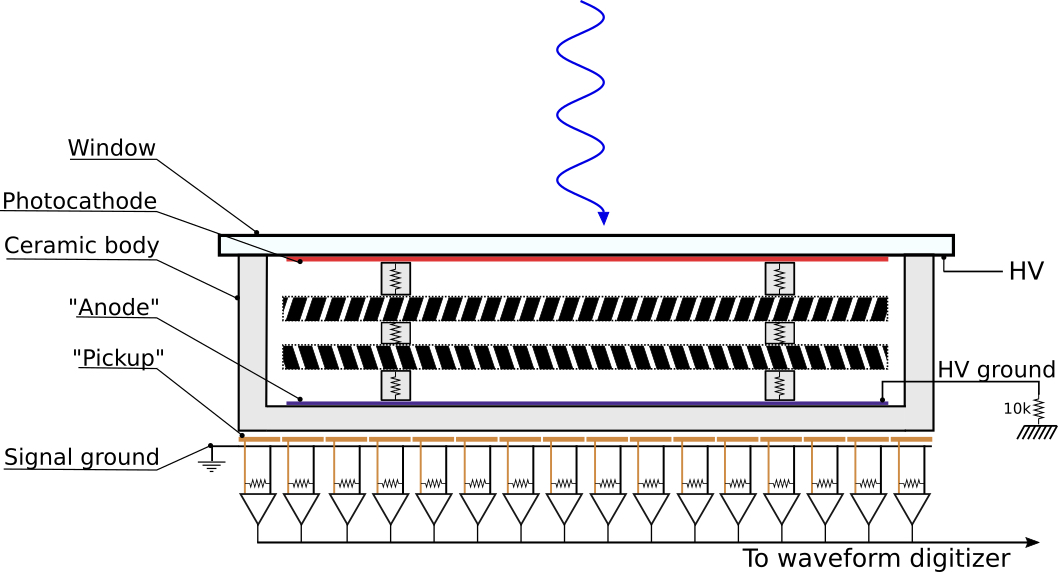}

\caption{The structure of the MCP-PMT with the capacitive coupling readout.
  The internal anode is a 10-nm-thick NiCr film, grounded through 2-10 k$\Omega$
  through a single connection from inside the vacuum package. The
  array of application-specific signal pickup electrodes outside of
  the vacuum package couples capacitively to the fast signal current
  in the thin metal anode film, and is measured across 50$\Omega$ relative
  to the signal ground at the digitization readout .}
\label{fig:MCP_sideview}
\end{figure}

The spatial and temporal resolutions are determined by the conversion
of the electron cascade pulse at the anode plane into a measurable
signal, and also by the transport of that signal through the package
wall to the digitizing electronics. Spatial resolutions of a few
microns can be obtained with anodes consisting of delay
lines~\cite{Ossy_JINST_2013,Ossy_delay_lines}, time resolutions of a
few 10's of picoseconds are measured using radio-frequency (RF) micro-strip delay
 lines~\cite{timing_paper,Tang_Naxos,anode_paper}. 
Two-dimensional arrays of pads allow a compromise, maintaining good resolution
in both space and time~\cite{Ohshima,Planacon}, but with many external
connections if the array is to have small pixels. 

In a capacitively coupled configuration, shown in
Fig.~\ref{fig:MCP_sideview}, the charge shower from the MCP
amplification section induces a current in a metallic anode deposited
on the bottom plate of the detector package. The signal pickup
consists of an array of conductors and a signal ground. The
pickup electrodes capacitively couple to currents on the internal
anode, producing a signal relative to the signal ground.

Traditionally, the anode readout pattern is sealed in the vacuum
packaging of the photodetector, and cannot be modified after
fabrication. In contrast, the capacitively coupled
readout pattern is separate from the detector, mounted outside of the
MCP-PMT vacuum packaging. The signal pickup can then be optimized for
different applications with different requirements on time resolution,
spatial resolution, and channel occupancy. The photodetector module and
capacitive pickup board can be manufactured independently.

Capacitively coupled MCP-PMT anodes have been successfully
demonstrated using resistive thick films or
semi-conductors~\cite{Jagutzki_1999, Jagutzki_2013, Photek_2014,
Jagutzki_2002, Photek_2007}\nocite{Battistoni_1978, Jagutzki_2007},
and MCP-PMTs using them are commercially
available~\cite{Photek_2014,RoentDek}.  Here we demonstrate that a
thin metal layer can also transmit the
high-frequency components of the fast pulses from an MCP-PMT to an
external array of electrodes.

The development and testing of the metal-film capacitively coupled anode
has been done in the context of the commercial development of
\twentysquare MCP-PMTs~\cite{Incom} following the R\&D of the LAPPD
Collaboration~\cite{history_paper}.  Metal film deposition is widely
available commercially, and making a uniform metal anode on the
interior of an \LAPPDTM module has proved to be much easier than
silk-screening and firing a resistive film.  In addition, since metal film
deposition is often already part of the process of photodetector
packaging, the deposition of an anode metal film can be easily integrated 
into the manufacturing.

\section{Anode and Signal Pickup Implementation}
\label{sec:anode_and_pickup}

\begin{figure}
\includegraphics[width = 0.56\columnwidth]{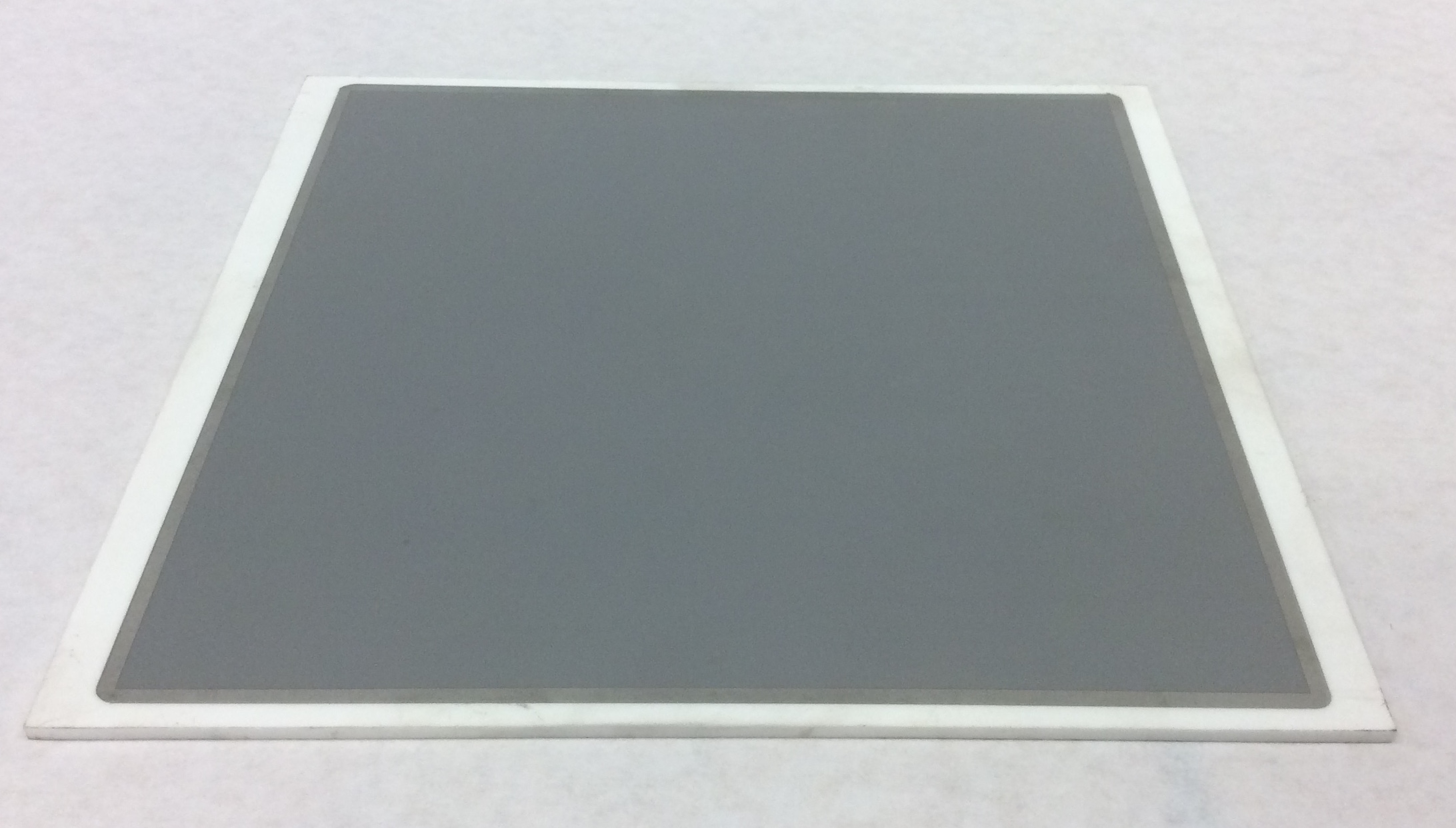}
\includegraphics[width=0.43\columnwidth]{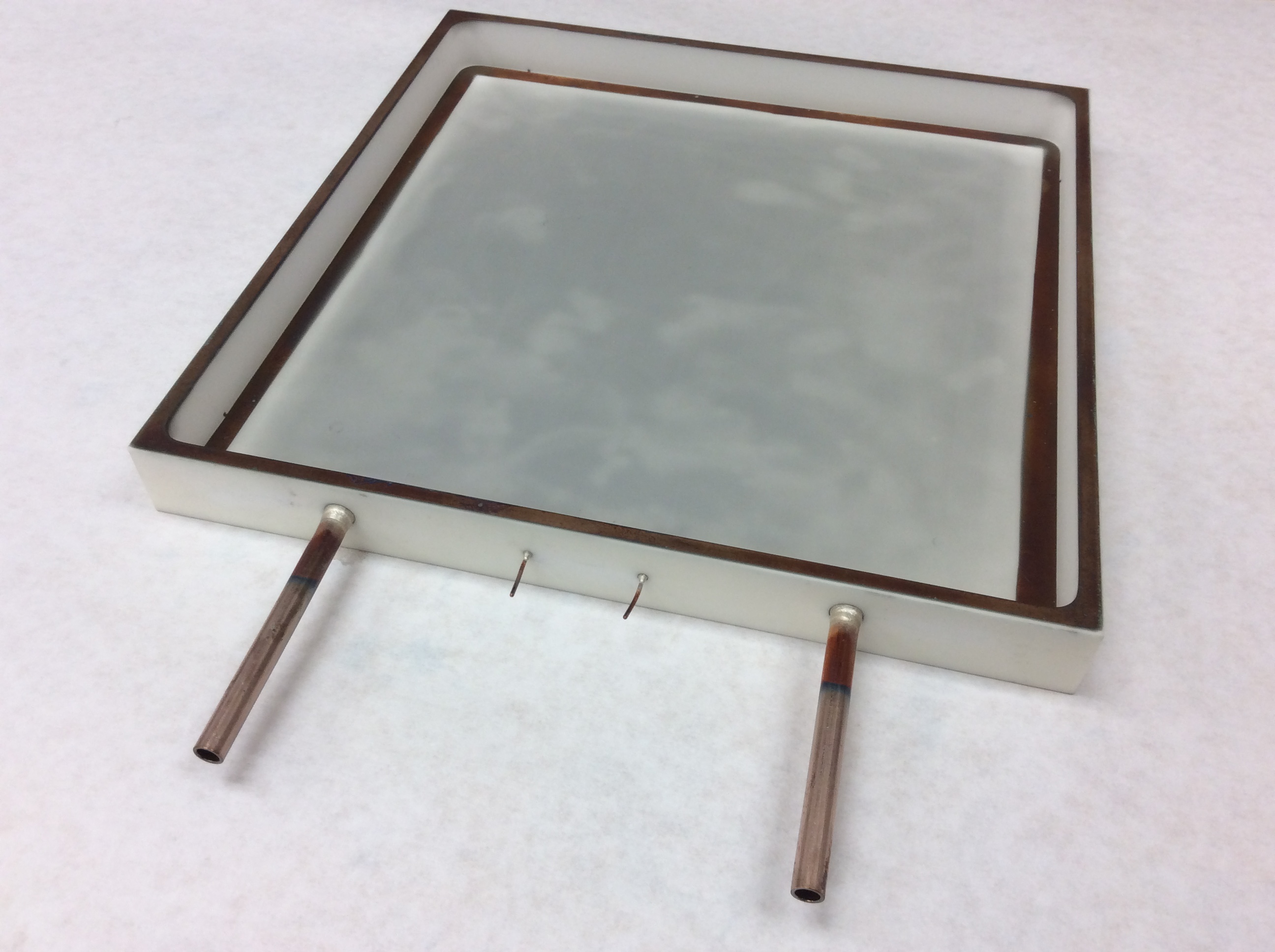}
\caption{Left: The ceramic anode plane with 10-nm-NiCr layer used in the present
test. Right: An LAPPD ceramic lower tile-assembly package with
evaporated 10-nm-NiCr anode. The  pins in the sidewall are used for the
anode connection to ground. The copper tubes are used for photo-cathode synthesis, and are not part of the electronics system.}
\label{fig:pickup_and_LTA}
\end{figure}

Testing has been done on a ceramic substrate that replicates
the bottom plate of a ceramic MCP-PMT package, shown in
Fig.~\ref{fig:pickup_and_LTA}.  The \twentysquare NiCr
metal anode is the same for the test substrate and for the MCP-PMT
package~\cite{thickness_difference}, and is described in
Section~\ref{sec:anode}.  The two pickup geometries, one of
two-dimensional pads and the other of micro-strips, are described in
Section~\ref{sec:pickup}.

%
%
\subsection{Ceramic/NiCr Anode Plane and Photodetector Package}
\label{sec:anode}
 
The test anode was constructed by evaporating a 10-nm-Nichrome
(NiCr)~\cite{NiCr} film onto a 3-mm-thick alumina-ceramic 
substrate~\cite{Coorstek_AD96R}.  In an operational MCP-PMT
the ceramic substrate acts as the bottom layer of the hermetic
package, with the anode on the inner surface.  The NiCr film of 
the anode has a sheet resistance of $\sim 100\Omega$ per square. 

\subsection{Signal Pickup Arrays}
\label{sec:pickup}

The geometric layout of readout conductors determines the temporal and
spatial resolutions of the photodetector. Typical anode patterns
include transmission-line micro-strips~\cite{anode_paper}, two-layer
crossed delay lines~\cite{Ossy_delay_lines,Jagutzki_1999}, and pad arrays
~\cite{Planacon}. 

\begin{figure}
\includegraphics[width = 1.0\columnwidth]{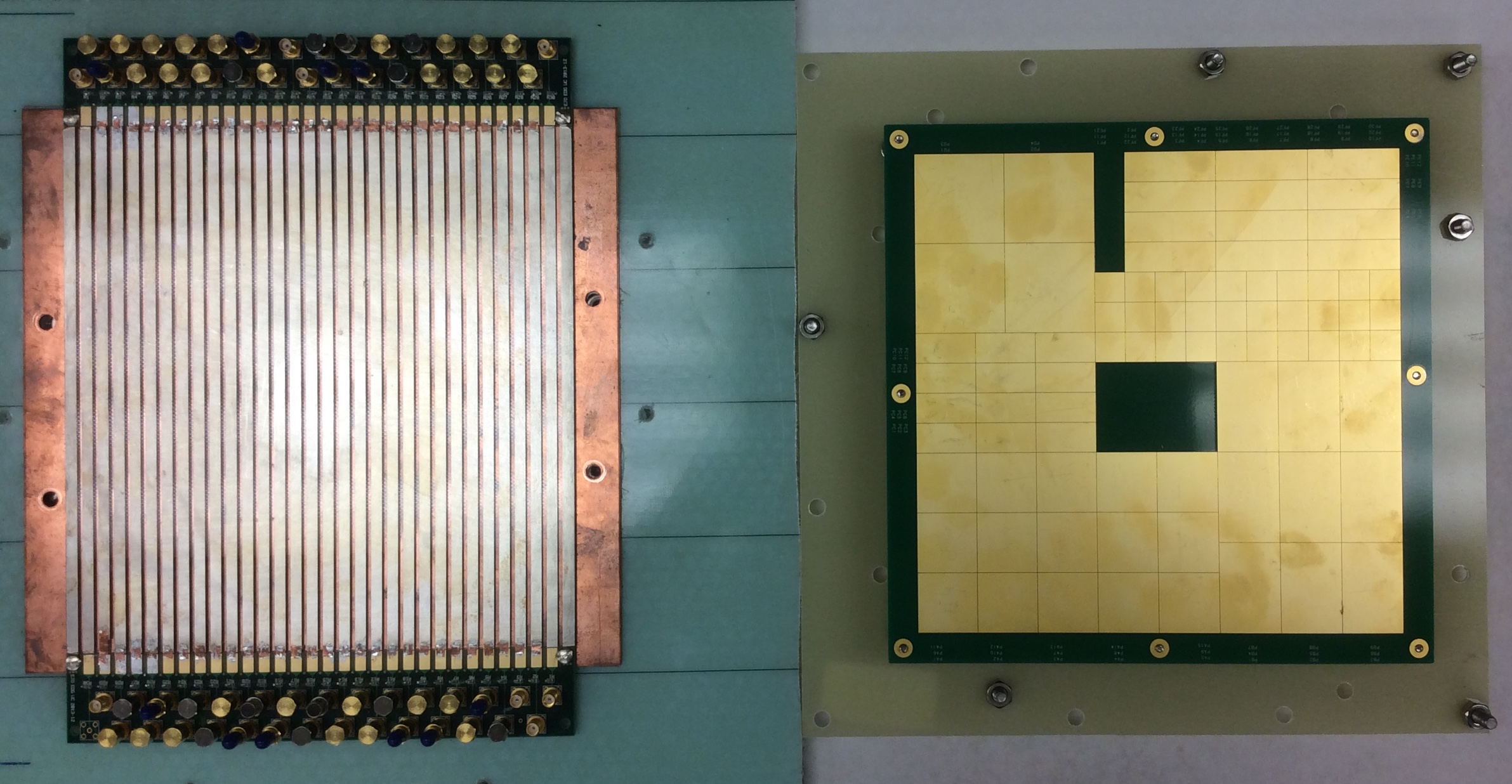}
\caption{The signal pickup planes for the micro-strip lines (left-hand
  panel) and 2-dimensional pad array (right-hand panel).}
\label{fig:strip_and_pad_pickups}
\end{figure}

Tests were performed using two separate signal pickups, one consisting
of 50$\Omega$ micro-strips, and the other an array of 2-dimensional pads.
The strip pickup, shown in the left-hand panel of
Figure~\ref{fig:strip_and_pad_pickups}, consists of silver strips
fired onto a glass substrate~\cite{anode_paper}. This pickup has an analog
bandwidth of 1.5 GHz and good signal characteristics up to a length of
90-cm~\cite{timing_paper,anode_paper}.

The capacitive coupling configuration enables using a pixelated
2-dimensional array of pads without brazed pins penetrating the
ceramic package.  A test pickup card, shown in the right-hand panel
of Figure~\ref{fig:strip_and_pad_pickups}, consists of a custom printed circuit
board with three sizes of square copper pads: 1.27 cm $\times$ 1.27 cm, 2.54
cm $\times$ 2.54 cm, and 3.81 cm $\times$ 3.81 cm.  The back side of
the board supplies signal ground through which each pad connects to an SMA
connector.

\section{Laser Test Stand}
\label{sec:laser_test_stand}

\begin{figure}[th]
\includegraphics[width = \columnwidth]{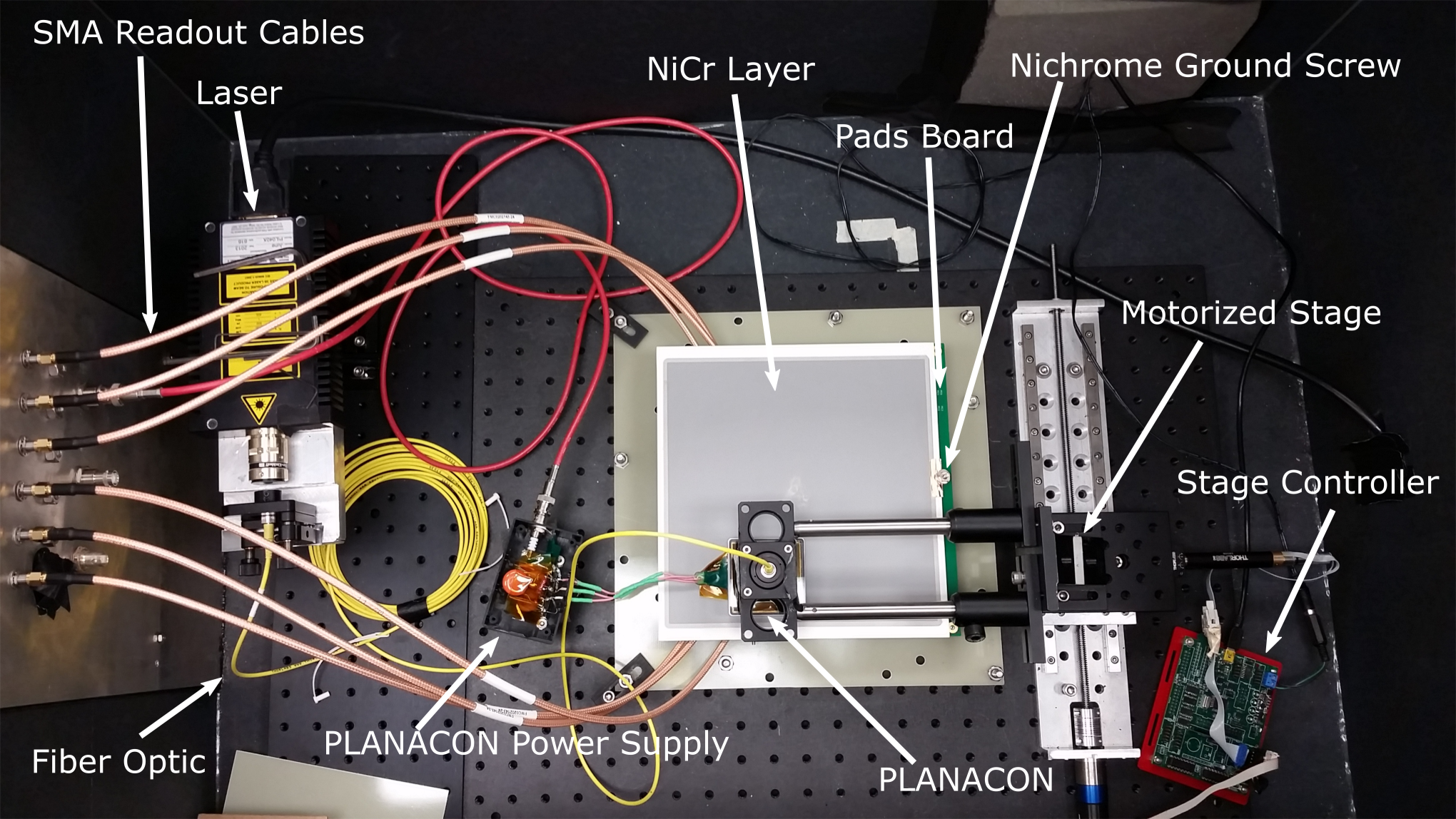}
\caption{The experimental setup for capacitive signal measurement. The
405 nm Pilas laser~\cite{laser} is fed through a fiber optic that is then
attached to a remotely-controlled translation stage. The laser is positioned
on the window of a Planacon MCP-PMT~\cite{Planacon}. The Planacon is coupled
 to the 10-nm-thick NiCr anode with uncured silver
epoxy. The signal pickup is directly beneath the anode
substrate.} 
\label{fig:setup_picture}
\end{figure}

Anodes were tested in both a `direct' configuration, in which one or
more individual conducting strips or pads of the signal pickup were
electrically connected to the signal source, and the capacitive
configuration, in which the signal source was connected to the NiCr
film anode and capacitively coupled to the pickup.  The
signal source consists of a commercial Planacon
MCP-PMT~\cite{Planacon}, illuminated by a 405 nm Pilas
laser~\cite{laser} fed through a fiber optic. The fiber is attached to a
remotely-controlled translation stage to the window, as shown in
Figure~\ref{fig:setup_picture}. The translation stage is equipped
with remote control PID motor drivers that allow for position scans
with a fiber position uncertainty of $<$ 1 mm~\cite{MotorDriver}. The
Planacon is coupled to the 10-nm-thick NiCr anode film evaporated on a
3-mm-thick ceramic substrate with uncured silver epoxy.

\subsection{Source of MCP generated pulses}
Previous tests of large-area microstrip performance have been
performed using MCP-PMT pulses generated by a Ti-Saph laser in a
pumped vacuum test facility~\cite{timing_paper, RSI_paper}, or on a
stand-alone anode in air with a fast pulse generator and a spectrum
analyzer~\cite{anode_paper}. Here, for convenience and compatibility with the dark box, a commercial Planacon MCP-PMT~\cite{Planacon} is used as the source of MCP pulses.

\subsection{Electrical Connections}

High voltage is distributed to the Planacon photocathode, the top,
and the bottom of the 2-MCP amplification stage  via
an external 50 M$\Omega$ resistive divider~\cite{Oberla_thesis}, with a
return from the anode to HV ground through 2k$\Omega$. The Planacon was
typically operated at a voltage of 1700 V.

The Planacon internal anode consists of an internal array of 32
$\times$ 32 metal pads that penetrate the vacuum packaging. An area of $\sim$ 13 cm$^2$ of the bottom of the Planacon was connected to the readout board using a consistent volume of uncured silver epoxy~\cite{silver_bond_50}. The epoxy provides a resistive path to ground for the collected charge, and a capacitive high-frequency connection to the signal readout. The resistance of the paste connection was small compared to the resistance of the HV
divider string so that the anode-DC offset was measured to be
negligible.

For the direct-coupling tests, the silver paste connects the
Planacon anode pads to  the signal
pickup board.  The signals are transmitted to the front-end
electronics via 50$\Omega$ cables.  

For the capacitive coupling tests, the Planacon anode pads are instead
connected via silver paste to the NiCr film on the top surface
of the ceramic substrate. The pickup board is located directly under
the ceramic substrate, as shown in Figure~\ref{fig:MCP_sideview}. The
signal connection to the front-end electronics is the same as for the
direct coupling case.

The effective impedance across which the signal current generates 
a voltage depends on the geometry of the pickup. The strip-line pickup
 has each end of the strips terminated
with 50$\Omega$, and there is substantial cross-talk between the
strips~\cite{anode_paper}. The effect of the 10-nm-Nicr anode 
is to spread the signal laterally among strips (see
Section~\ref{sec:crosstalk}); the effective impedance to signal ground
 is thus substantially lower than 50$\Omega$.

\subsection{Readout Electronics and Data Acquisition}

The readout electronics system was based on the PSEC4 fast waveform
digitization Application Specific Integrated Circuit (ASIC), a custom,
6-channel waveform digitizer that samples at 10-15
GSamples/sec~\cite{PSEC4_paper}. The PSEC4 is available as a
stand-alone single-chip system with FPGA 
local control, read out by a laptop computer via USB. 
 A calibrated PSEC4 waveform digitizer has an
analog bandwidth of 1.5 GHz and a typical RMS noise level of 700
$\mu$V~\cite{PSEC4_paper}. The tests in this paper, done on an
uncalibrated chip in a noisy environment, have a measured noise level
of 2.2 mV, to be compared with the $\sim$ 50 - 300mV of the  
typical Planacon signals.

%
%

\section{Results}
\label{sec:results}

\subsection{Rise-Time and Amplitude Comparison}
\label{sec:risetime_and_amplitude}

\begin{figure}[th]
\includegraphics[width = 0.49\columnwidth]{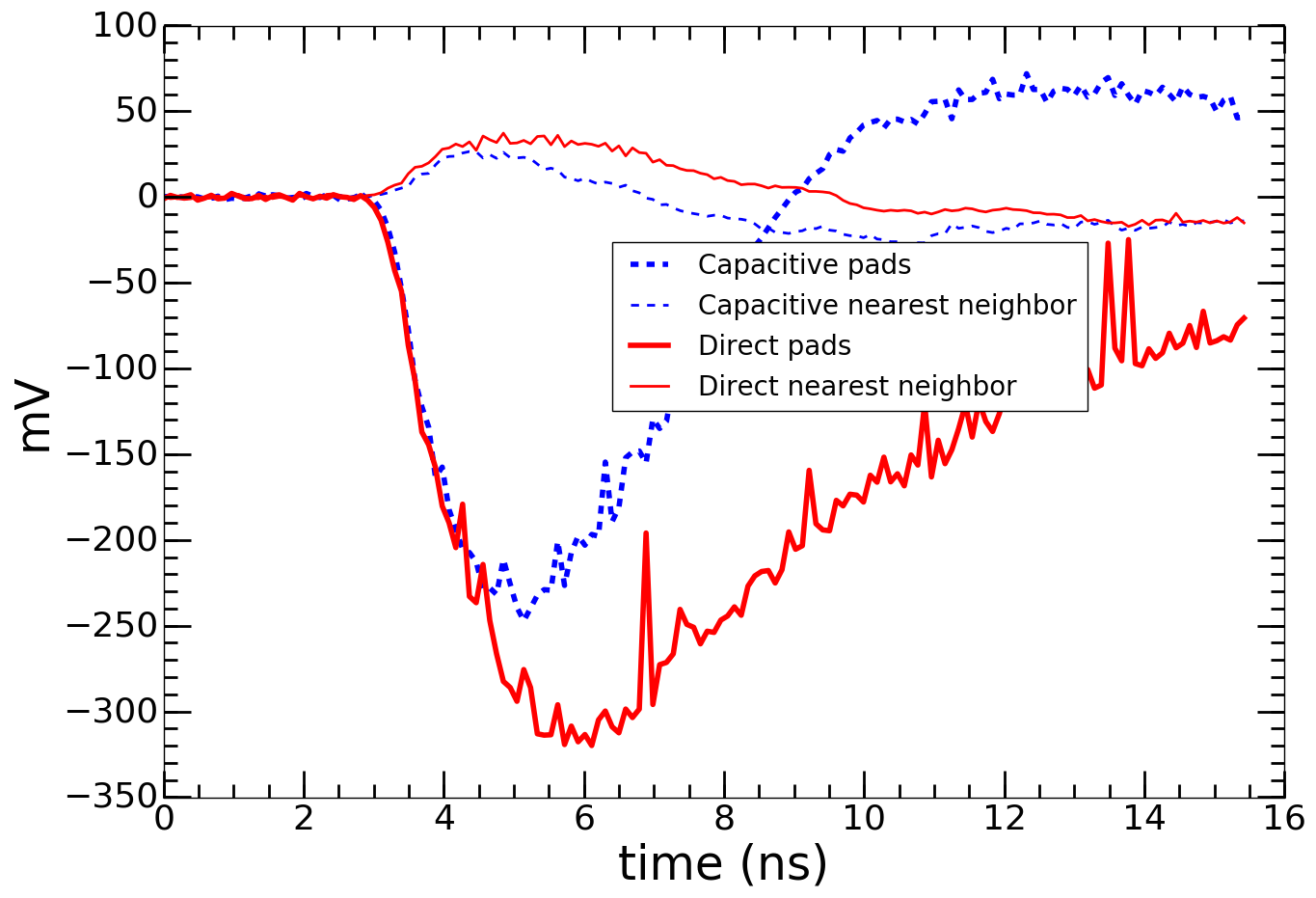}
\hfil
\includegraphics[width = 0.49\columnwidth]{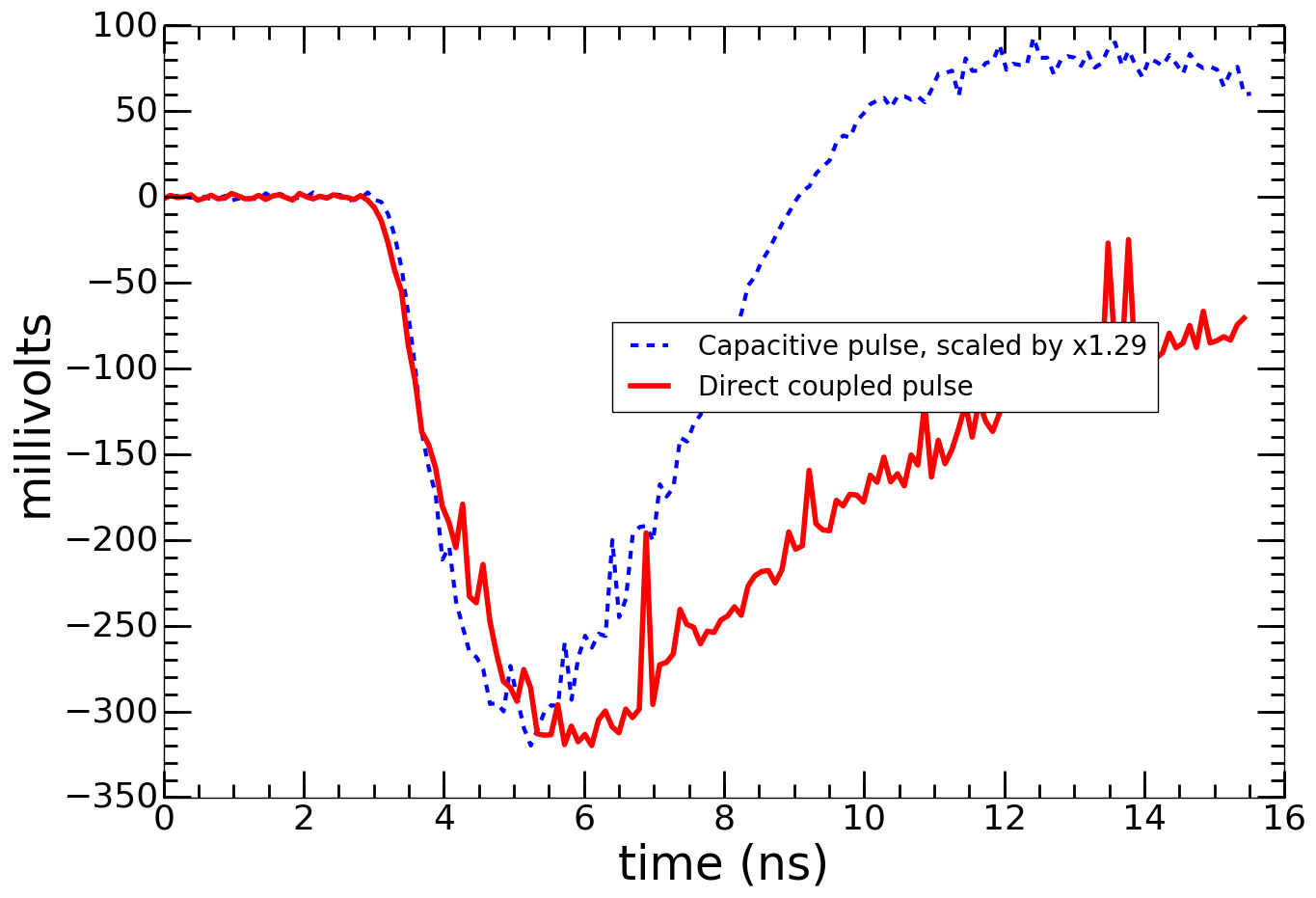}
\includegraphics[width = 0.49\columnwidth]{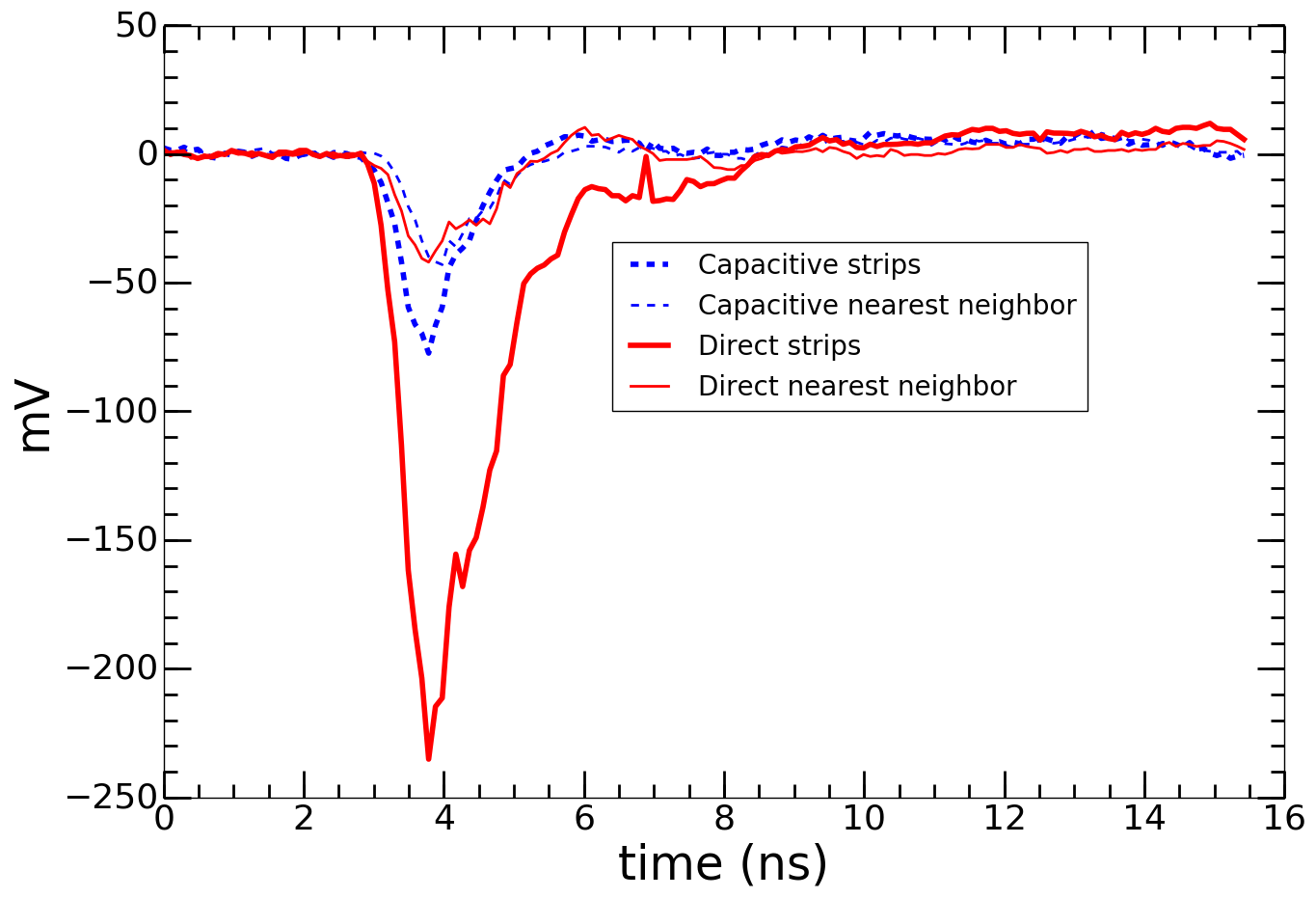}
\hfil
\includegraphics[width = 0.49\columnwidth]{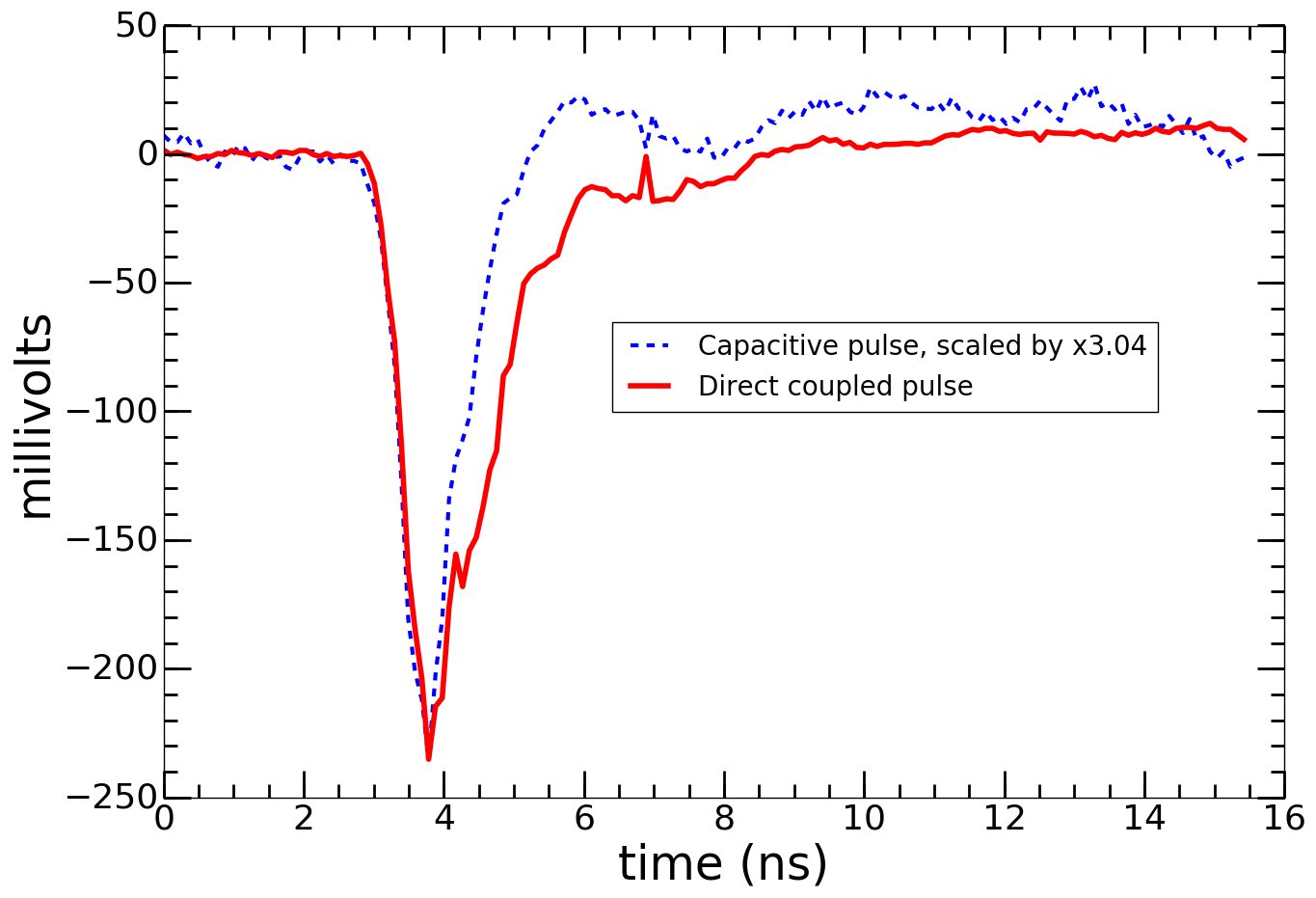}

\caption{Left: A typical pulse trace on the (red) direct coupling and (blue)
capacitive coupling configurations for \oneinchsquare pads (upper
panels) and 0.44 cm-wide micro-strips on a 0.69 cm pitch (lower panels). Two
traces are shown for each configuration: the pad or strip directly
beneath the laser head and a neighboring pad or
strip. Right: The pulse traces normalized to the same amplitude,
showing that the high-frequency response is unchanged by the
capacitive coupling through the metal film.}
\label{fig:strip_pads_overlays}
\end{figure}

Figure~\ref{fig:strip_pads_overlays} shows typical individual pulses
recorded by the test system. The panels in the upper row show a direct
pulse (red solid line) and a capacitively coupled pulse (blue, dashed
line) for the \oneinchsquare pad configuration. The right-hand panel
shows the capacitively coupled pulses normalized to the largest direct
pulse. Approximately 80\% of the direct amplitude 
is picked up by one pad in the capacitive geometry, 
with no degradation in risetime.

The two panels in the lower row show the same comparison for the
micro-strip pickup configuration. In the capacitive case, shown in the
left-hand panel, 34\% of the direct amplitude is measured on the
strip directly under the laser, and substantial portions of the amplitude are recorded
on neighboring strips. The right-hand panel shows the direct
and capacitive pulses normalized by amplitude; the rise times and
pulse widths are essentially identical.  The comparison between the
top and bottom rows shows the inherently higher bandwidth of the RF
micro-strips versus the two-dimensional geometry of the pads.

\subsection{Cross-Talk and Position resolution}
The physical placement of the ceramic substrate between the
NiCr anode and the pickup electrodes spreads the signal,
and thus changes the spatial resolution and coupling between
neighboring pickup electrodes. Each electrode is separately
digitized so in principle no signal is lost as long as each
signal is large compared to the single-channel noise. The `cross-talk'
between neighboring electrodes allows for spatial resolution
better than the strip pitch. However, cross-talk can present problems
for pattern recognition in high-occupancy applications~\cite{disambiguation_paper}.

%
%

\subsubsection{Micro-strip  pickup}
The micro-strip signal pickup used here is identical in construction to
that of Ref.~\cite{anode_paper}, in which the anode was excited
directly by LAPPD MCPs.  Crosstalk between neighboring micro-strips was
used to measure a spatial resolution of 700 \micron by
interpolation~\cite{anode_paper}.

\begin{figure}[th]
\centering
\includegraphics[width = 0.65\columnwidth]{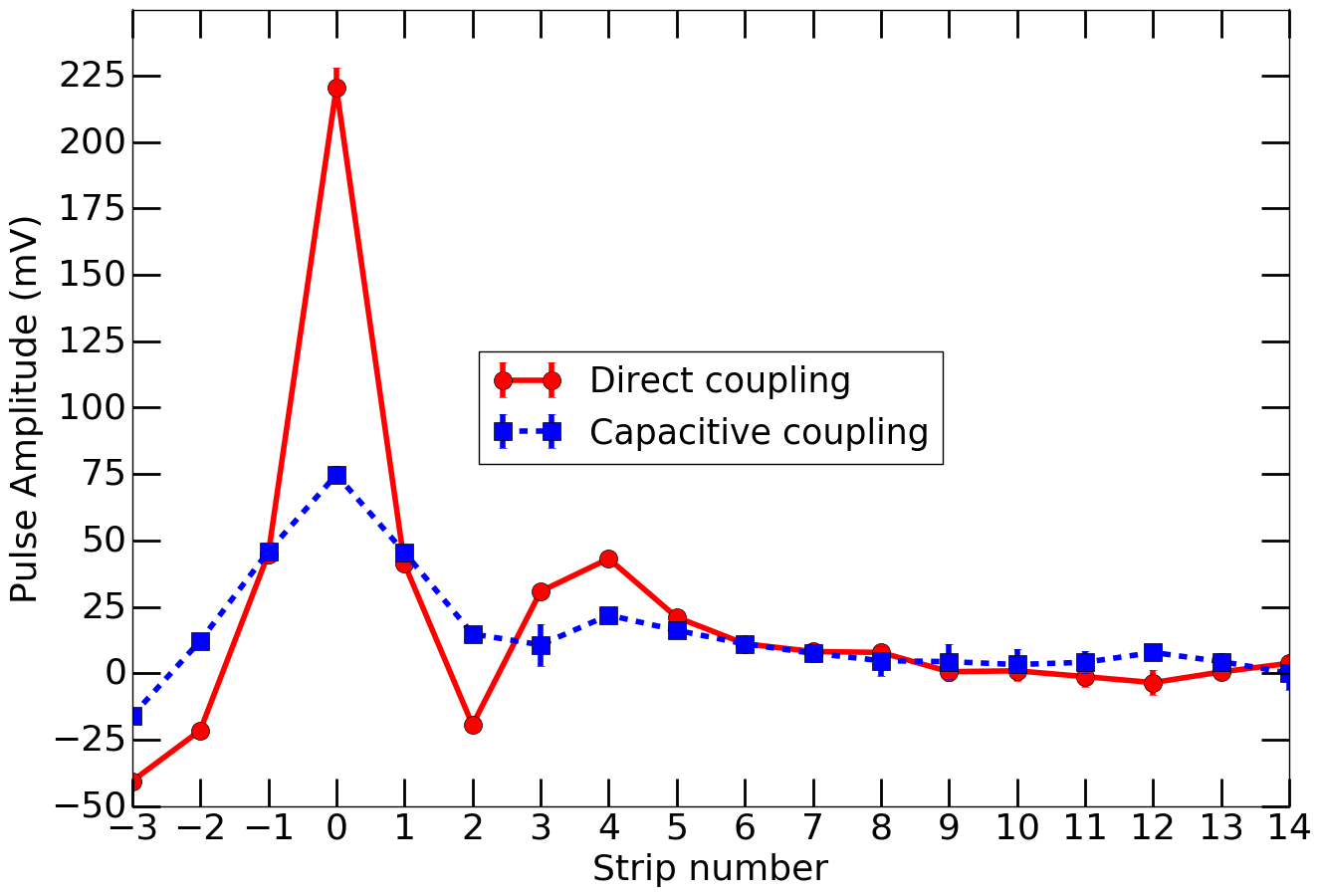}
\caption{Measurement of the signal amplitude on each pickup strip. The
strips are spaced center-to-center by 6.9 mm.  The signal spreads out
along the NiCr anode, and forms a lower impedance capacitive coupling
 to more strips in the pickup layer beneath.}
\label{pulsedistribution}
\end{figure}

A comparison of the distributions in pulse amplitude on each strip for
the directly coupled (red solid line) and capacitively coupled (blue
dashed line) configurations is shown in Figure \ref{pulsedistribution}.
Relative to the directly coupled configuration, the signal is
spread to neighboring strips, and the amplitude on the central
strip-line is $\sim$66\% smaller.  We attribute this effect to the
intervening extended NiCr film, distributing the current laterally, and
forming a lower impedance coupling to the pickup strips below.

\label{sec:crosstalk}

\subsubsection{Pad-array pickup}
The optimal size of the pads in the signal pickup is
application-specific, and depends on the desired spatial resolution
and expected occupancy. For example, a large neutrino detector deep
underground will be characterized by low rate and typically low
occupancy, with a resolution that depends on size and analysis
technique~\cite{doublebeta_paper1}. In contrast, a
detector at a high-energy hadron collider will typically have very
high occupancy, and a resolution that depends on the specific use,
such as time-of-flight for charged particles~\cite{Credo} or
calorimetry~\cite{CMS_calorimetry}. The ease of changing the pad size
allows accommodation of varying rate or resolution requirements even in
a single application, for example changing the pad size as a function
of rapidity at a collider detector.

%
%

\begin{figure}[th]
\includegraphics[width = 0.49\columnwidth]{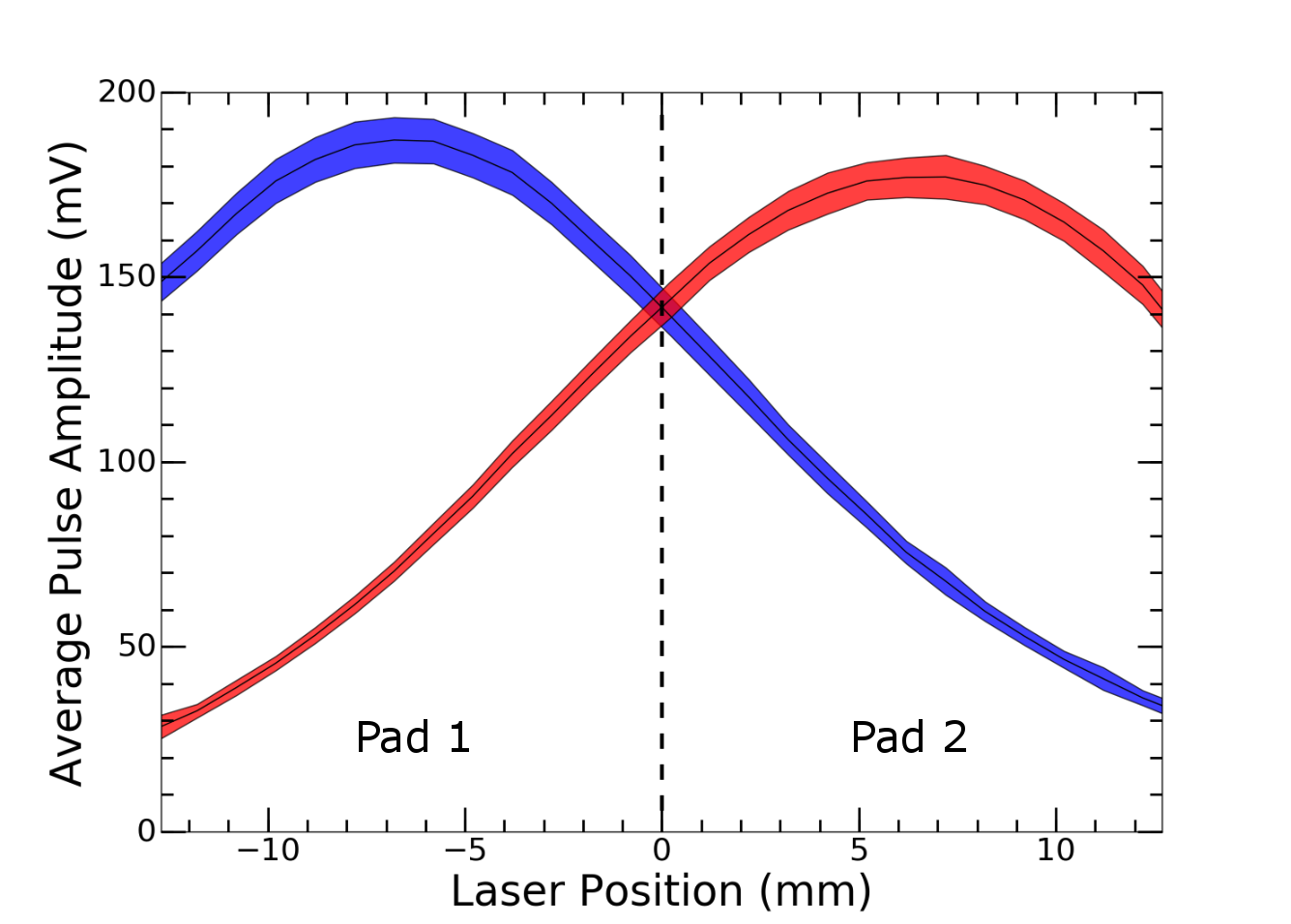}
\includegraphics[width = 0.49\columnwidth]{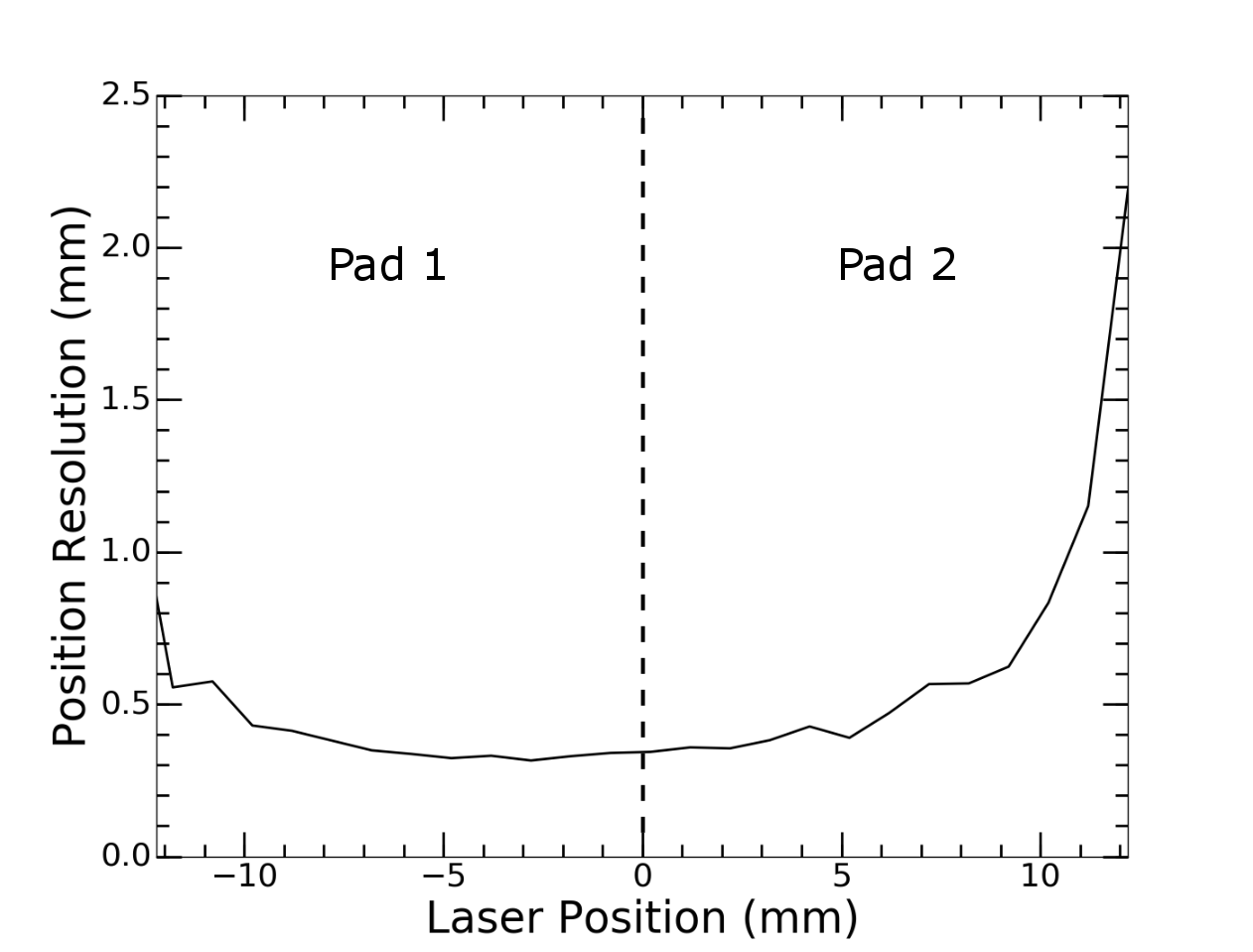}
\caption{Left: The average amplitude as a function of laser position on
half inch pads. The shaded region is $\pm 1\sigma$  in each
direction. Right: The position resolution as the laser is scanned across two
adjacent 1.27 cm pads.}\label{PositionScanandResolution}
\end{figure}

The test pad signal pickup board was constructed with pads of
different sizes, motivated by the range needed for high-luminosity
collider applications. The results on position resolution are
measured by scanning the laser fiber position over a 30 mm range along
an array of 1.27 cm pads.

The left-hand panel of Figure \ref{PositionScanandResolution} shows
the signal amplitude on two 1.27 cm pads as a function of laser position as
the laser is scanned across their interface. The right-hand panel
shows the position resolution calculated from the fraction
$f = A_1/(A_1+A_2)$, where $A_1$ is the amplitude on the first pad,
and $A_2$ is the amplitude on the second pad.  The resolution is
slowly-varying as long as the laser signal is shared between the two
pads, sharply rising when only one pad contains the signal, and
has a minimum of 315 $\mu$m when the laser is centered on the
interface.




\section{Conclusions}

We have tested a  \twentysquare thin-metal-film
internal anode capacitively coupled to an array of signal pads or
micro-strips for use in fast \mcppmts. 
The pads or micro-strips constituting the signal pickup are located on a 
plane that can be put in close proximity to the bottom
outer surface of the sealed photodetector, with no electrical
connections through the photodetector hermetic vacuum package other
than a single ground connection to the internal anode.

The tests were performed with a Planacon MCP-PMT excited by a Pilas
laser as the signal source.  For the capacitively coupled
configuration, the Planacon was coupled electrically to the test anode
plane, which consisted of a 10-nm-thick NiCr film deposited on a 96-percent-pure
 Al$_2$O$_3$ 3-mm-thick ceramic plate. The test anode was connected
to HV ground through 2k$\Omega$. On the lower side of the ceramic plate,
signals were picked up by a plane with signal electrodes on the top
surface and a continuous signal ground on the bottom surface. Two
signal-pickup readout patterns were tested: 1) parallel 50$\Omega$
25-cm-long micro-strips with an analog bandwidth of 1.5 GHz, and 2) a
\twentysquare array of 2-dimensional square pads with sides of 1.27 cm or 2.54 cm. Signals were digitized with a PSEC4 10 GSa/s waveform
sampling system.

The results were compared with those obtained with the Planacon
directly coupled to the signal pickup planes. The
capacitive coupling has no effect on the measured rise-time of the
pulse for both the pad and micro-strip pickup geometries, preserving
the analog bandwidth for timing resolution. For capacitively coupled
pads of \oneinchsquare, we observe 80\% of the directly coupled
amplitude.  For the micro-strip pattern, we measure 34\% of the
directly coupled signal on the strip with the largest signal,
still large enough for precision spatial measurements. We attribute
the loss on the excited strip to a lower impedance to signal ground
due to crosstalk with multiple micro-strips, each terminated at both
ends in 50$\Omega$.

Cross-talk between neighboring electrodes on the signal pickup card
allows spatial resolution more precise than the size of the individual
electrodes. Scanning with a translation stage, we find that on \halfinchsquare pads, position resolution is nearly constant over a region in which there is signal sharing between 2 neighboring pads, with an optimal resolution of 315 $\mu$m.

\section{Acknowledgments}
The activities at the University of Chicago were supported by the
Department of Energy under DE-SC-0008172 and the National Science
Foundation under grant PHY-1066014.  We owe many thanks to E. Oberla
for help with the PSEC4-based electronics readout. We thank F. Tang (UC Electronics Design Group) for
designing the pad signal pickup card; C. Craven, M. Foley, M. J.
Minot (Incom, Inc) and B. Shi (Thin Film Deposition Lab, Argonne
APS) for the deposition of thin metal films on the
ceramic anode planes; J. DeFazio (Photonis, Inc) for support of the
Planacon photodetector; and R. Northrop (UC Engineering Center) and
M. Heintz (UC EDG) for essential mechanical engineering and
computational support.



\linespread{1}\selectfont


\begin{thebibliography}{99}
\bibitem{thickness_difference}
 Due to availability of ceramic plates in stock the test
anode is 3-mm-thick; the base of the MCP-PMT is 5mm thick.
\end{thebibliography}
\end{document}